\documentclass[12pt, letterpaper]{article}

\newtheorem{define}{D{\small EFINITION}}
\newtheorem{lemma}{L{\small EMMA}}
\newtheorem{theorem}{T{\small HEOREM}}
\newtheorem{assumption}{A{\small SSUMPTION}}

\usepackage{natbib}
\usepackage{amsmath}
\usepackage{amsfonts}
\usepackage{amssymb}
\usepackage{latexsym}
\usepackage{setspace}
\usepackage{authblk}

\usepackage{graphicx}
\usepackage{subcaption}
\usepackage{varwidth}

\RequirePackage[l2tabu, orthodox]{nag}

\begin{document}
\title{\begin{spacing}{1}Optimal Bandwidth Selection for the Fuzzy Regression Discontinuity Estimator\end{spacing}}
\author[a]{Yoichi Arai\thanks{Corresponding Author.}}
\author[b]{Hidehiko Ichimura}
\affil[a]{\small National Graduate Institute for Policy Studies (GRIPS), 7-22-1 Roppongi, Minato-ku, Tokyo 106-8677, Japan; yarai@grips.ac.jp}
\affil[b]{Department of Economics, University of Tokyo, 7-3-1 Hongo, Bunkyo-ku, Tokyo, 113-0033, Japan; ichimura@e.u-tokyo.ac.jp}
\date{}
\maketitle

\begin{abstract}
\begin{spacing}{1.2}
A new bandwidth selection method for the fuzzy regression
 discontinuity estimator is proposed.
The method chooses two bandwidths simultaneously, one for each side of
 the cut-off point by using a criterion based on the
 estimated asymptotic mean square error taking into account a
 second-order bias term. A simulation study demonstrates the usefulness
 of the proposed method.
\end{spacing}
\end{abstract}
\begin{spacing}{1.2}
\end{spacing}

\newpage
\section{Introduction}
The fuzzy regression discontinuity (FRD) estimator, developed by
\citet{htv01} (hereafter HTV), has found
numerous empirical applications in economics.
The target parameter in the FRD design is the ratio of the difference of two conditional mean functions, which is interpreted as the local average treatment effect.
The most frequently used estimation method is the nonparametric method using the local linear regression (LLR).
\citet{ik12} (hereafter IK) propose a bandwidth selection method specifically aimed at the FRD estimator, which uses a single bandwidth to estimate all conditional mean functions.

This paper proposes to choose two bandwidths simultaneously, one for each side of the cut-off point.
In the context of the sharp RD (SRD) design, \citet{ai15} (hereafter AI) show that the
simultaneous selection method is theoretically superior to the existing
methods and their extensive simulation experiments verify the
theoretical predictions.
We extend their approach to the FRD estimator.
A simulation study illustrates the potential usefulness of the
proposed method.\footnote{Matlab and Stata codes to implement
the proposed method are available at \texttt{http://www3.grips.ac.jp/\~{}yarai/}.}

\section{Bandwidth Selection of The FRD Estimator}\label{sec:fuzzy}
For individual $i$ potential outcomes with and without
treatment are denoted by $Y_{i}(1)$ and $Y_{i}(0)$, respectively.
Let $D_{i}$ be a binary variable that stands for the treatment status, 0 or 1.
Then the observed outcome, $Y_{i}$, is described as $Y_{i} = D_{i}Y_{i}(1) + (1-D_{i}) Y_{i}(0)$.
Throughout the paper, we assume that $(Y_{1}, D_{1}, X_{1})$, $\ldots$, $(Y_{n}, D_{n}, X_{n})$ are i.i.d. observations and $X_{i}$ has the Lebesgue density $f$.

To define the parameter of interest for the FRD design, denote $m_{Y+}(x) = E(Y_{i}|X_{i}=x)$ and $m_{D+}(x) = E(D_{i}|X_{i} = x)$ for $x\geq c$.
Suppose that $\lim_{x \searrow c} m_{Y+}(x)$ and $\lim_{x \searrow c} m_{D+}(x)$ exist and they are denoted by $m_{Y+}(c)$ and $m_{D+}(c)$, respectively.
We define $m_{Y-}(c)$ and $m_{D-}(c)$ similarly.
The conditional variances and covariance, $\sigma_{Yj}^{2}(c)>0$, $\sigma_{Dj}^{2}(c)>0$, $\sigma_{YDj}(c)$, and the second and third derivatives $m_{Yj}^{(2)}(c)$, $m_{Yj}^{(3)}(c)$, $m_{Dj}^{(2)}(c)$, $m_{Dj}^{(3)}(c)$, for $j=+, -$, are defined in the same manner.
We assume all the limits exist and are bounded above.

In the FRD design, the treatment status depends on the assignment variable $X_{i}$ in a stochastic manner and the propensity score function is known to have a discontinuity at the cut-off point $c$, implying $m_{D+}(c) \ne m_{D-}(c)$.
Under the conditions of HTV, \citet{por03} or \citet{dl14}, the LATE at the cut-off point is given by $\tau (c) = (m_{Y+}(c) - m_{Y-}(c))/(m_{D+}(c) - m_{D-}(c))$.
This implies that estimation of $\tau(c)$ reduces to estimating the four conditional mean functions nonparametrically and the most popular method is the LLR because of its automatic boundary adaptive property (\citealp{fa92}). 

Estimating the four conditional expectations, in principle, requires four bandwidths.
IK simplifies the choice by using a single bandwidth to estimate all
functions as they do for the SRD design.
For the SRD design, AI proposes to choose bandwidths, one
for each side of the cut-off point because the curvatures of the
conditional mean functions and the sample sizes on the left and the right of the cut-off
point may differ significantly.
We use the same idea here, but take into account the bias and variance
due to estimation of the denominator as well.
For simplification, we propose to choose one bandwidth, $h_{+}$, to estimate $m_{Y+}(c)$ and
$m_{D+}(c)$ and another bandwidth, $h_{-}$, to estimate $m_{Y-}(c)$ and
$m_{D-}(c)$ because it is also reasonable to use the same group on each side.

\subsection{Optimal Bandwidths Selection for the FRD Estimator}
We consider the estimator of $\tau(c)$, denoted $\hat\tau(c)$, based on the LLR estimators of
the four unknown conditional mean functions.
We propose to choose two bandwidths simultaneously based on an asymptotic approximation of the mean squared error (AMSE).
To obtain the AMSE, we assume the following:
\begin{assumption}\label{assumption:kernel}
(i) (Kernel) $K(\cdot): \mathbb{R}\to\mathbb{R}$ is a symmetric second-order kernel function that is continuous with compact support; 
(ii) (Bandwidth) The positive sequence of bandwidths is such that $h_{j}\to 0$ and $nh_{j}\to \infty$ as $n\to \infty$ for $j=+, -$.
\end{assumption}

Let ${\mathcal D}$ be an open set in $\mathbb{R}$, $k$ be a nonnegative integer, ${\mathcal C}_{k}$ be the family of $k$ times continuously differentiable functions on ${\mathcal D}$ and $g^{(k)}(\cdot)$ be the $k$th derivative of $g(\cdot)\in {\mathcal C}_{k} $.
Let ${\mathcal G}_{k}({\mathcal D})$ be the collection of functions $g$ such that $g\in {\mathcal C}_{k}$ and $\left| g^{(k)}(x) - g^{(k)}(y)\right|\le M_{k}\left|x-y\right|^{\alpha}$, $x, y, z \in {\mathcal D}$,
for some positive $M_{k}$ and some $\alpha$ such that $0<\alpha \le 1$.

\begin{assumption}\label{assumption:density}
The density of $X$, $f$, which is bounded above and strictly positive at
 $c$, is an element of ${\mathcal G}_{1}({\mathcal D})$ where ${\mathcal D}$ is an open neighborhood of $c$.
\end{assumption}
\begin{assumption}\label{assumption:function}
Let $\delta$ be some positive constant. 
The $m_{Y+}$, $\sigma_{Y+}^{2}$ and $\sigma_{YD+}$ are elements of ${\mathcal G}_{3}({\mathcal D}_{1})$, ${\mathcal G}_{0}({\mathcal D}_{1})$ and ${\mathcal G}_{0}({\mathcal D}_{1})$, respectively, where ${\mathcal D}_{1}$ is a one-sided open neighborhood  of $c$, $(c, c+\delta)$.
Analogous conditions hold for $m_{Y-}$, $\sigma_{Y-}^{2}$ and $\sigma_{YD-}$ on ${\mathcal D}_{0}$ where ${\mathcal D}_{0}$ is a one-sided open neighborhood of $c$, $(c-\delta, c)$.
\end{assumption}

The following approximation holds for the MSE under the conditions stated above.
\begin{lemma}\label{lemma:MSE2F}
Suppose Assumptions \ref{assumption:kernel}--\ref{assumption:function} hold.
Then, it follows that
\begin{multline}
 MSE_{n}(h_{+},h_{-}) =\frac1{(\tau_{D}(c))^2} \left\{ \Bigl[\phi_1(c) h_{+}^{2} - \phi_0(c) h_{-}^{2}\Bigr] + \Bigl[\psi_1(c) h_{+}^{3} - \psi_0(c) h_{-}^{3}\Bigr] + o\left(h_{+}^3 + h_{-}^3 \right) \right\}^{2}\\
 + \frac{v}{n f(c) (\tau_{D}(c))^{2}} \left\{ \frac{\omega_1(c)}{h_{+}} + \frac{\omega_0(c)}{h_{-}}  \right\} + o\left(\frac1{nh_{+}}+\frac1{nh_{-}}\right)
\end{multline}
where, for $j=+, -$ and $k = Y, D$, $\tau_{D}(c) = m_{D+}(c) - m_{D-}(c)$, $\omega_j(c)  = \sigma_{Yj}^{2}(c) +\tau(c)^{2} \sigma_{Dj}^{2}(c) -2 \tau(c) \sigma_{YDj}(c)$, $\phi_{j}(c) = C_1 \left[m_{Yj}^{(2)}(c)  - \tau(c) m_{Dj}^{(2)}(c)\right]$, $\psi_j(c) = \zeta_{Yj}(c) - \tau(c)\zeta_{Dj}(c)$,
\begin{align*}
\zeta_{kj}(c)  &=  (-j) \left\{\xi_{1}\left[ \frac{m_{kj}^{(2)}(c)}{2} \frac{f^{(1)}(c)}{f(c)} + \frac{m_{kj}^{(3)}(c)}{6} \right]  -  \xi_{2} \frac{ m_{kj}^{(2)}(c)}{2} \frac{f^{(1)}(c)}{f(c)} \right\},
\end{align*}
$C_{1} = \left(\mu_{2}^{2} - \mu_{1}\mu_{3})/2(\mu_{0} \mu_{2} - \mu_{1}^{2} \right)$,
$v = (\mu_{2}^{2}\nu_{0} - 2 \mu_{1}\mu_{2}\nu_{1} + \mu_{1}^{2}\nu_{2})/(\mu_{0}\mu_{2}-\mu_{1}^2)^{2}$,
$\xi_{1}  = (\mu_{2}\mu_{3} - \mu_{1}\mu_{4})/(\mu_{0}\mu_{2} - \mu_{1}^{2})$,
$\xi_{2}= (\mu_{2}^{2} - \mu_{1}\mu_{3}) \left(\mu_{0}\mu_{3} - \mu_{1}\mu_{2}\right)/( \mu_{0}\mu_{2} - \mu_{1}^{2})^{2}$,
$\mu_{j} = \int_0^{\infty} u^{j} K(u) du$, $\nu_{j} = \int_0^{\infty} u^{j} K^{2}(u) du$.
\end{lemma}

A standard approach applied in this context is to minimize the following AMSE, ignoring higher order terms:
\begin{equation}\label{eq:AMSE}
AMSE(h_{+},h_{-}) =\frac1{(\tau_{D}(c))^{2}} \Bigl\{ \phi_{+}(c) h_{+}^{2} - \phi_{-}(c) h_{-}^{2} \Bigr\}^{2} + \frac{v}{n f(c) (\tau_{D}(c))^{2}} \left\{ \frac{\omega_1(c)}{h_{+}} + \frac{\omega_0(c)}{h_{-}}  \right\}.
\end{equation}
As AI observed, 
(i) while the optimal bandwidths that minimize the AMSE (\ref{eq:AMSE})
are well-defined when $\phi_{+}(c)\cdot \phi_{-}(c)<0$, they are not
well-defined when $\phi_{+}(c)\cdot \phi_{-}(c)>0$ because the bias term
can be removed by a suitable choice of bandwidths and the bias-variance
trade-off breaks down.\footnote{This is the reason why IK proceed with assuming $h_{+} = h_{-}$.}
(ii) When the trade-off breaks down, a new optimality criterion becomes
necessary in order to take higher-order bias terms into consideration.
We define the asymptotically first-order optimal (AFO)
bandwidths, following AI.
\begin{define}\label{def:AFOF}
The AFO bandwidths for the FRD estimator \emph{minimize the AMSE defined by
\[
{AMSE}_{1n}(h_{+},h_{-})= \frac1{(\tau_{D}(c))^2} \Bigl\{ \phi_{+}(c)h_{+}^2 - \phi_{-}(c)h_{-}^2 \Bigr\}^2 + \frac{v}{n f(c) (\tau_{D}(c))^2} \left\{\frac{\omega_{+}(c)}{h_{+}}+ \frac{\omega_{-}(c)}{h_{-}}\right\},
\]
when $\phi_{+}(c) \cdot \phi_{-}(c)<0$. 
When} $\phi_{+}(c) \cdot \phi_{-}(c)>0$, the AFO bandwidths for the FRD estimator \emph{minimize the AMSE defined by
\[
AMSE_{2n}(h_{+},h_{-})= \frac1{(\tau_{D}(c))^2} \Bigl\{ \psi_{+}(c) h_1^3 - \psi_{-}(c)h_{-}^3  \Bigr\}^2 + \frac{v}{n f(c) (\tau_{D}(c))^2} \left\{\frac{\omega_{+}(c)}{h_{+}}+ \frac{\omega_{-}(c)}{h_{-}}\right\}
\]
subject to the restriction $\phi_{+}(c)h_{+}^2 - \phi_{-}(c)h_{-}^2=0$ under the assumption of $\psi_{+}(c) -  \{\phi_{+}(c)/\phi_{-}(c)\}^{3/2} \psi_{-}(c)\ne 0$.
}
\end{define}
When $\phi_{+}(c)\cdot \phi_{-}(c)<0$, the AFO bandwidths minimize the standard AMSE (\ref{eq:AMSE}).
When $\phi_{+}(c)\cdot \phi_{-}(c)>0$, the AFO bandwidths minimize the
sum of the squared second-order bias term and the variance term under
the restriction that the first-order bias term be removed.
Inspecting the objective function, the resulting AFO bandwidths are $O(n^{-1/5})$ when $\phi_{+}(c)\cdot \phi_{-}(c)<0$ and $O(n^{-1/7})$ when $\phi_{+}(c)\cdot \phi_{-}(c)>0$.\footnote{The explicit expression of the AFO bandwidths are provided in the Supplemental Material.}

When $\phi_{+}(c) \cdot \phi_{-}(c)>0$,
Definition 1 shows that $MSE_{n}$ is of order $O(n^{-6/7})$, which
implies that the MSE based on the AFO bandwidths converges to zero
faster than $O(n^{-4/5})$, the rate attained by the single bandwidth
approaches such as the IK method.
When $\phi_{+}(c) \cdot \phi_{-}(c)<0$, they are of the same order.
However, it can be shown that the ratio of the AMSE based on the AFO
bandwidths to that based on IK never exceeds one asymptotically (see Section 2.2 of AI).

\subsection{Feasible Automatic Bandwidth Choice}\label{sec:feasible}
The feasible bandwidths is based on a modified version of the estimated
AMSE (MMSE) as in AI.  It is defined by
\begin{multline}\label{eq:MMSEF}
{MMSE}_{n}^p(h_{+},h_{-}) = \Bigl\{ \hat \phi_{+}(c) h_1^2 - \hat \phi_{-}(c) h_{-}^2 \Bigr\}^{2} +  \Bigl\{ \hat \psi_{+}(c) h_1^3 - \hat \psi_{-}(c) h_{-}^3 \Bigr\}^2\\
 + \frac{v}{n \hat f(c)} \left\{\frac{\hat \omega_{+}(c)}{h_1}+ \frac{\hat \omega_{-}(c)}{h_{-}}\right\}
\end{multline}
where $\hat \phi_{j}(c)$, $\hat \psi_{j}(c)$, $\hat \omega_{j}(c)$ and $\hat f(c)$  are consistent estimators of $\phi_{j}(c)$, $\psi_{j}(c)$, $\omega_{j}(c)$ and $f(x)$ for $j=+, -$, respectively.
A key characteristic of the MMSE is that one does not need to know the
sign of the product of the second derivatives a priori and that there is
no need to solve the constrained minimization problem.

Let $(\hat h_{+}, \hat h_{-})$ be a combination of bandwidths that minimizes the MMSE given in (\ref{eq:MMSEF}).
The next theorem shows that $(\hat h_{+}, \hat h_{-})$ is asymptotically as good as the AFO bandwidths.
\begin{theorem}\label{theorem:MMSEF}
Suppose that the conditions stated in Lemma \ref{lemma:MSE2F} hold. 
Assume further that  $\hat \phi_{j}(c) \stackrel{p}{\to} \phi_{j}(c)$, $\hat \psi_{j}(c) \stackrel{p}{\to} \psi_{j}(c)$, $\hat f(c)\stackrel{p}{\to} f(c)$  and $\hat \omega_{j}(c)\stackrel{p}{\to} \omega_{j}(c)$ for $j=+, -$, respectively.
Also assume $\psi_{+}(c) -  \{\phi_{+}(c)/\phi_{-}(c)\}^{3/2} \psi_{-}(c)\ne 0$.
Then, the following hold.
\[
\frac{\hat h_{+}}{h_{+}^{\dagger}} \stackrel{p}{\to}  1, \quad \frac{\hat h_{-}}{h_{-}^{\dagger}} \stackrel{p}{\to} 1, \quad \mbox{and} \quad \frac{{MMSE}_{n}^p(\hat h_{+},\hat h_{-})}{MSE_{n}(h_{+}^{\dagger}, h_{-}^{\dagger})} \stackrel{p}{\to} 1.
\]
where $(h_{+}^{\dagger}, h_{-}^{\dagger})$ are the AFO bandwidths.
\end{theorem}
The first part of Theorem \ref{theorem:MMSEF} shows that the bandwidths based on the plug-in version of the MMSE are asymptotically equivalent to the AFO bandwidths and the second part exhibits that the minimum value of the MMSE is asymptotically the same as the MSE evaluated at the AFO bandwidths.
Theorem \ref{theorem:MMSEF} shows that the bandwidths based on the MMSE possess the desired asymptotic properties. 
Theorem \ref{theorem:MMSEF} calls for pilot estimates for $\phi_{j}(c)$, $\psi_{j}(c)$, $f(c)$ and $\omega_{j}(c)$ for $j=+, -$.
A detailed procedure about how to obtain the pilot estimates is given in the Supplemental Material.

\section{Simulation}
We conduct simulation experiments that illustrate
the advantage of the proposed method and
a potential gain for using bandwidths
tailored to the FRD design over the bandwidths tailored to the SRD design.
Application of a bandwidth developed for the SRD to the FRD context seems common in practice.\footnote{For example, \citet[Section 5.1]{ik12} state that the bandwidth choice for the FRD estimator is often similar to the choice for the SRD estimator of only the numerator of the FRD estimand.}

Simulation designs are as follows.
For the treatment probability, $E[D|X=x]=\frac1{\sqrt{2\pi}}\int_{-\infty}^{x}\exp[-\frac{(u+1.28)^{2}}{2}]du$ for $x\geq 0$ and $\frac1{\sqrt{2\pi}}\int_{-\infty}^{x}\exp[-\frac{(u-1.28)^{2}}{2}]du$ for $x<0$.
This leads to the discontinuity size of 0.8.
The graph is depicted in Figure 1-(a).

For the conditional expectation functions of the observed outcome,
 $E[Y|X=x]$, we consider two designs, which are essentially the same as Designs 2 and 4 of AI for the the SRD design.
The specification for the assignment variable and the additive error are exactly the same as that considered by IK.\footnote{The exact functional form of $E[Y|X=x]$, the specification for the assignment variable and the additive error are provided in the Supplemental Material}
They are depicted in Figures 1-(b) and 1-(c).
We use data sets of 500 observations with 10,000 repetitions.

\begin{figure}[htbp]
    \begin{subfigure}[b]{0.33\linewidth}
    \centering
    \captionsetup{justification=centering}
    \includegraphics[width=1\linewidth]{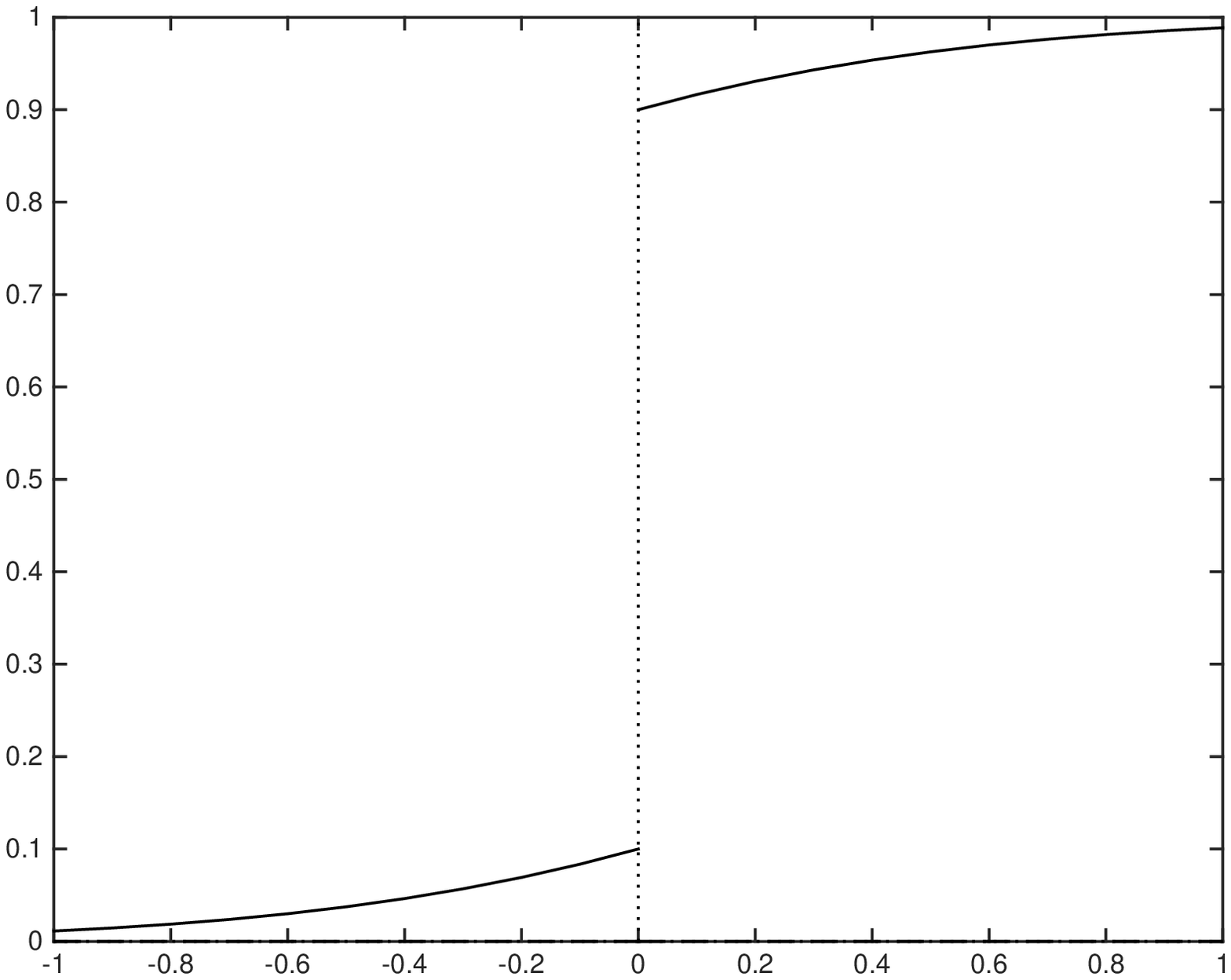} 
  \caption{Treatment Probability, $E[D|X=x]$.}
  \label{figure:px} 
  \end{subfigure}
  \begin{subfigure}[b]{0.33\linewidth}
    \centering
    \captionsetup{justification=centering}
    \includegraphics[width=1\linewidth]{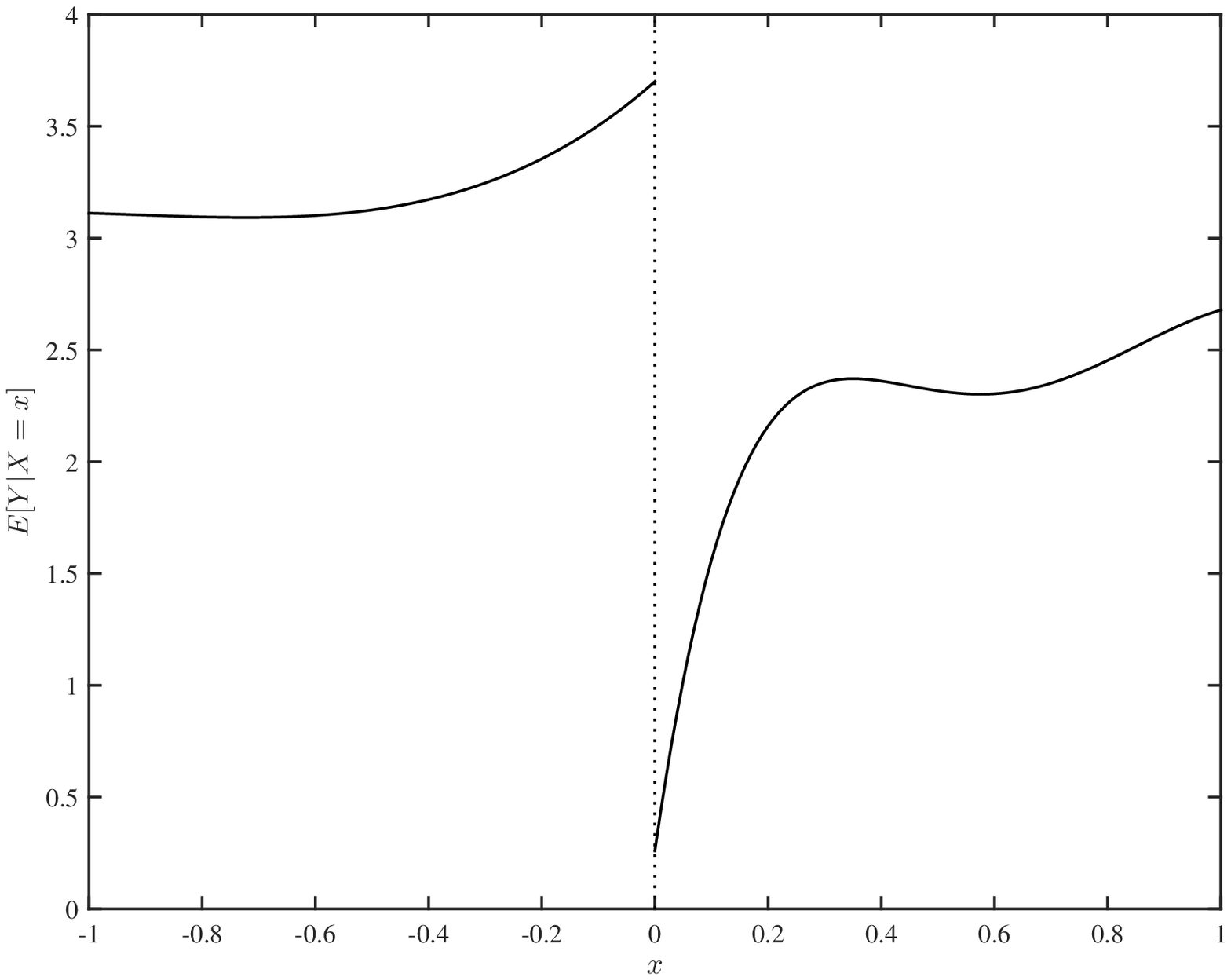} 
    \captionsetup{singlelinecheck=off}
    \caption[.]{Design 1. Ludwig and Miller Data I}
    \label{figure:dgp:a} 
  \end{subfigure}
  \begin{subfigure}[b]{0.33\linewidth}
    \centering
    \captionsetup{justification=centering}
    \includegraphics[width=1\linewidth]{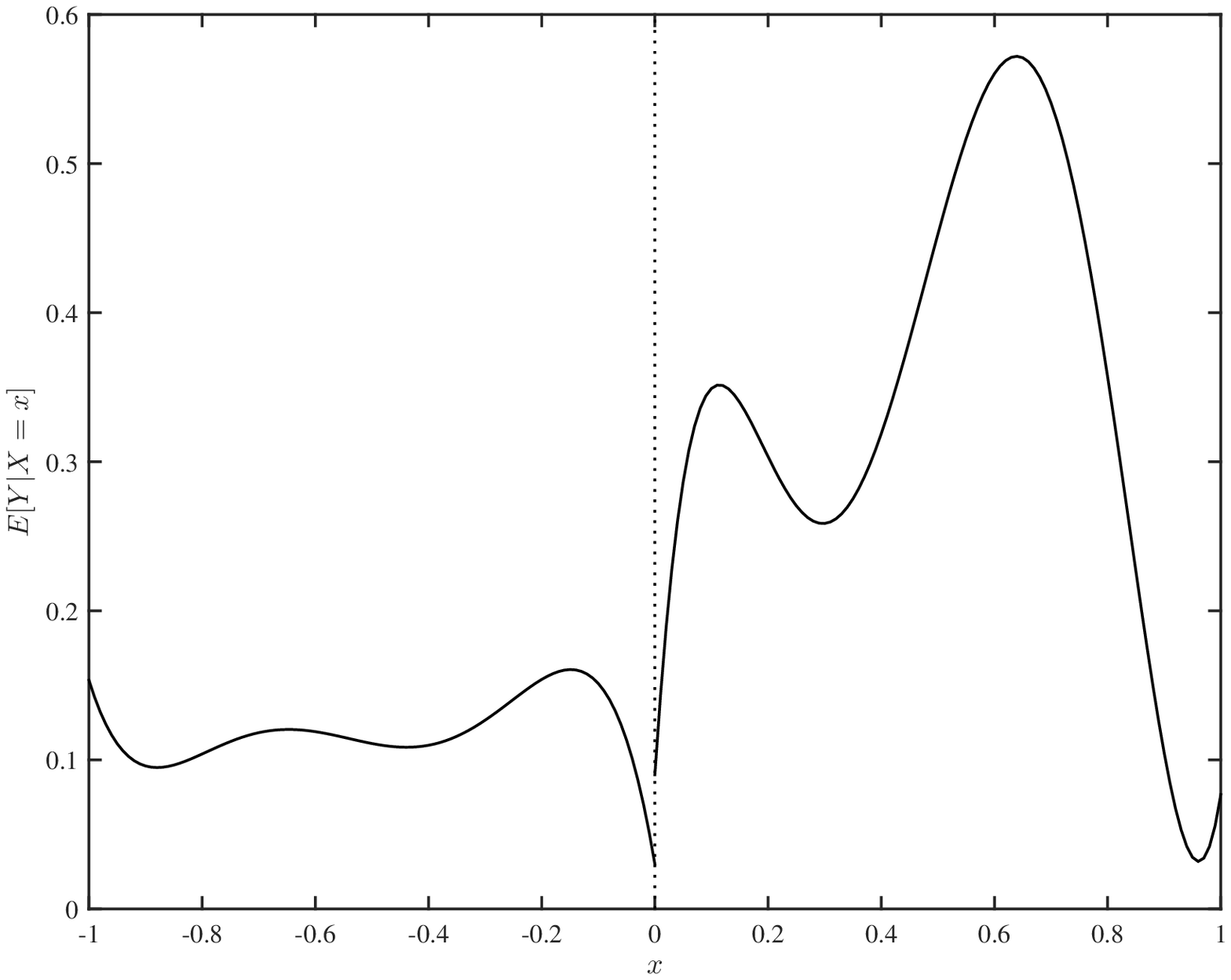} 
    \captionsetup{singlelinecheck=off}
    \caption[]
  {Design 2. Ludwig and Miller Data II}
    \label{figure:dgp:b} 
  \end{subfigure} 
  \caption{Simulation Designs. (a) Treatment probability, (b)  $E[Y|X=x]$ for Design 1, (c) $E[Y|X=x]$ for Design 2.}
  \label{figure:dgp} 
\end{figure}

\begin{figure}[htbp]
    \begin{subfigure}[b]{0.5\linewidth}
    \centering
    \captionsetup{justification=centering}
    \includegraphics[width=1\linewidth]{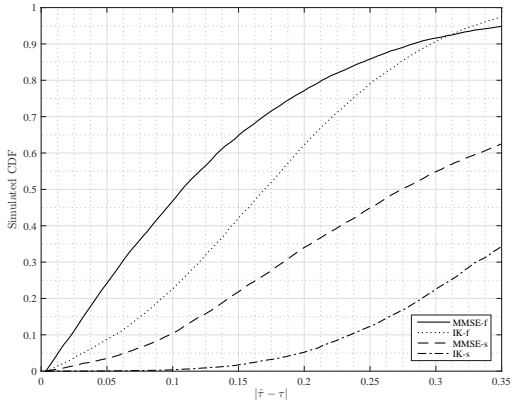} 
  \caption{Design 1. Ludwig and Miller Data I}
  \label{figure:scdf1} 
  \end{subfigure}
  \begin{subfigure}[b]{0.5\linewidth}
    \centering
    \captionsetup{justification=centering}
    \includegraphics[width=1\linewidth]{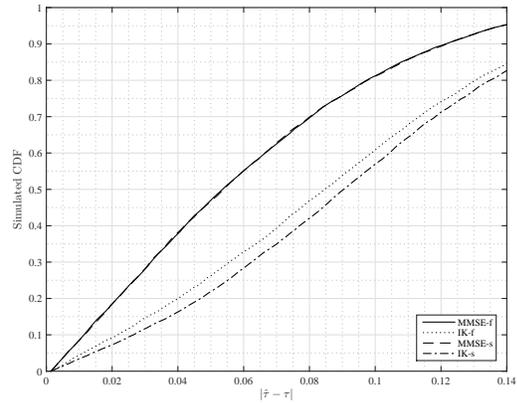} 
    \captionsetup{singlelinecheck=off}
    \caption[.]{Design 2. Ludwig and Miller Data II}
    \label{figure:scdf2} 
  \end{subfigure}
  \caption{Simulated CDF of $|\hat\tau - \tau|$ for different bandwidth selection rules}
  \label{figure:scdf} 
\end{figure}

The results are presented in Table \ref{table:frd02} and Figure \ref{figure:scdf}.
Four bandwidth selection methods, MMSE-f, MMSE-s, IK-f and IK-s, are examined.
MMSE-f is the new method proposed in the paper and MMSE-s is the one
proposed for the SRD design by AI.
IK-f and IK-s are the methods proposed by IK for the FRD and SRD designs, respectively.
The bandwidths for MMSE-s and IK-s are computed based only on the numerator of the FRD estimator and the same bandwidths are used to estimate the denominator.
Table \ref{table:frd02} reports the mean and standard deviation of the bandwidths, the bias and root mean squared error (RMSE) for the FRD estimates, and the relative efficiency based on the RMSE.\footnote{The bias and RMSE are 5\% trimmed versions since unconditional finite sample variance is infinite.}
Figure 2 shows the simulated CDF for the distance of the FRD estimate from the true value.

Examining Table 1 and Figure 2-(a), for Design 1, MMSE-f performs
significantly better than all other methods.
For Design 2, Table 1 and Figure 2-(b) indicate that MMSE-f and MMSE-s performs
comparably but clearly dominate IK methods currently widely used.
In cases we examined, the new method performs better than currently
available methods and using methods specifically developed for the FRD dominates the method
developed for the SRD.

\begin{table}
\caption{Bias and RMSE for the FRDD, n=500}
\label{table:frd02}
\begin{center}
\small
\begin{tabular}{llccccccc}
\hline \hline
& & \multicolumn{2}{c}{$\hat h_{+}$} & \multicolumn{2}{c}{$\hat h_{-}$}  & \multicolumn{3}{c}{$\hat \tau$}\\ 
Design & Method & Mean & SD & Mean & SD & Bias & RMSE & Efficiency \\ \hline
1 & MMSE-f & 0.056 & 0.033 & 0.097 & 0.100 & 0.037 & 0.168 & 1 \\
 & IK-f & 0.177 & 0.033 &  &  & 0.180 & 0.192 & 0.876 \\
 & MMSE-s & 0.235 & 0.109 & 0.489 & 0.259 & 0.339 & 0.373 & 0.450 \\
 & IK-s & 0.325 & 0.068 &  &  & 0.443 & 0.451 & 0.373 \\ \hline
2 & MMSE-f & 0.226 & 0.091 & 0.624 & 0.214 & -0.002 & 0.072 & 1 \\
 & IK-f & 0.284 & 0.052 &  &  & 0.087 & 0.097 & 0.739 \\
 & MMSE-s & 0.227 & 0.092 & 0.628 & 0.215 & -0.002 & 0.072 & 1 \\
 & IK-s & 0.337 & 0.072 &  &  & 0.093 & 0.101 & 0.709 \\
\hline
\end{tabular}
\end{center}
\end{table}

\section*{Acknowledgement}
This research was supported by Grants-in-Aid for Scientific Research No. 22243020 and No. 23330070 from the Japan Society for the Promotion of Science. Yoko Sakai provided expert research assistance.

\bibliography{yarai}
\bibliographystyle{econometrica}

\newpage
\appendix

\noindent {\bf \LARGE Supplemental Material}\\ \\
This Supplemental Material provides an explicit expression of the AFO bandwidths, details about our simulation experiment, a sketch of the proof for Lemma 1 and procedures to obtain pilot estimates.

\section{Definition of the AFO Bandwidths}

\begin{define}\label{def:AFOF}
The AFO bandwidths for the fuzzy RD estimator \emph{minimize the AMSE defined by
\[
{AMSE}_{1n}(h)= \frac1{(\tau_{D}(c))^2} \Bigl\{ \phi_{+}(c)h_{+}^2 - \phi_{-}(c)h_{-}^2 \Bigr\}^2 + \frac{v}{n f(c) (\tau_{D}(c))^2} \left\{\frac{\omega_{+}(c)}{h_{+}}+ \frac{\omega_{-}(c)}{h_{-}}\right\}.
\]
when $\phi_{+}(c) \cdot \phi_{-}(c)<0$. 
Their explicit expressions are given by $h_{+}^{*} =  \theta^{*} n^{-1/5}$ and $h_{-}^{*} = \lambda^{*} h_{+}^{*}$, where
\begin{equation}\label{eq:lambdastar}
\theta^{*} = \left\{ \frac{v \omega_{+}(c)}{4 f(c) \phi_{+}(c) \left[ \phi_{+}(c) - {\lambda^{*}}^{2} \phi_{-}(c) \right]} \right\}^{1/5} \quad \mbox{and} \quad \lambda^{*} = \left\{- \frac{\phi_{+}(c)\omega_{-}(c)}{\phi_{-}(c)\omega_{+}(c)} \right\}^{1/3}.
\end{equation}
When} $\phi_{+}(c) \cdot \phi_{-}(c)>0$, the AFO bandwidths for the fuzzy RD estimator \emph{minimize the AMSE defined by
\[
AMSE_{2n}(h)= \frac1{(\tau_{D}(c))^2} \Bigl\{ \psi_{+}(c) h_1^3 - \psi_{-}(c)h_{-}^3  \Bigr\}^2 + \frac{v}{n f(c) (\tau_{D}(c))^2} \left\{\frac{\omega_{+}(c)}{h_{+}}+ \frac{\omega_{-}(c)}{h_{-}}\right\}
\]
subject to the restriction $\phi_{+}(c)h_{+}^2 - \phi_{-}(c)h_{-}^2=0$ under the assumption of $\psi_{+}(c) -  \{\phi_{+}(c)/\phi_{-}(c)\}^{3/2} \psi_{-}(c)\ne 0$.
Their explicit expressions are given by $h_{+}^{**} = \theta^{**} n^{-1/7}$ and $h_{-}^{**} = \lambda^{**} h_{+}^{**}$, where
\[
\theta^{**} = \left\{ \frac{v \left[\omega_{+}(c) + \omega_{-}(c)/\lambda^{**} \right] }{6 f(c) \left[ \psi_{+}(c) - {\lambda^{**}}^{3} \psi_{-}(c) \right]^{2}}  \right\}^{1/7} \quad \mbox{and}\quad \lambda^{**} = \left\{\frac{\phi_{+}(c)}{\phi_{-}(c)} \right\}^{1/2}.
\]
}
\end{define}

\section{Simulation Designs}
Let $\ell_{1}(x) = E[Y(1)|X=x]$ and $\ell_{0} = E[Y(0)|X=x]$.
A functional form for each design is given as follows:
\begin{itemize}
\item[(a)] Design 1
\begin{align*}
\ell_{j}(x) &= \left\{
\begin{array}{ll}
\alpha_{j} + 18.49x - 54.8x^2 + 74.3x^3 - 45.02x^4 + 9.83x^5 &\mbox{ if } x >0,\\
\alpha_{j} + 2.99x + 3.28x^2 + 1.45x^3 + 0.22x^4 + 0.03x^5 & \mbox{ if } x \le 0,
\end{array}
\right.
\end{align*}
where $(\alpha_{1},\alpha_{0}) = (-0.17, 4.13)$.
\item[(b)] Design 2
\begin{align*}
\ell_{j}(x) &= \left\{
\begin{array}{ll}
\alpha_{j} + 5.76 x - 42.56 x^2 + 120.90 x^3 - 139.71x^4 + 55.59 x^5 &\mbox{ if } x >0,\\
\alpha_{j} - 2.26 x - 13.14 x^2 -30.89 x^3  -31.98 x^4 -12.1 x^5 & \mbox{ if } x \le 0,
\end{array}
\right. 
\end{align*}
where $(\alpha_{1},\alpha_{0}) = (0.0975, 0.0225)$.
\end{itemize}

The assignment variable $X_{i}$ is given by $2Z_{i}-1$ for each design where $Z_{i}$ have a Beta distribution with parameters $\alpha=2$ and $\beta=4$.
We consider a normally distributed additive error term with mean zero and standard deviation $0.1295$ for the outcome equation.

\section{Proofs}
\noindent\textbf{Proof of Lemma 1}:  
As in the proof of Lemma A2 of Calonico, Cattaneo, and Titiunik (2014), we utilize the following expansion
\begin{multline}\label{eq:expansion}
\frac{\hat\tau_{Y}(c)}{\hat\tau_{D}(c)}  - \frac{\tau_{Y}(c)}{\tau_{D}(c)} = \frac1{\tau_{D}(c)} \left(\hat\tau_{Y}(c) - \tau_{Y}(c) \right) - \frac{\tau(c)}{\tau_{D}(c)} \left( \hat\tau_{D}(c) - \tau_{D}(c) \right)\\
 + \frac{\tau(c)}{\tau_{D}(c) \hat\tau_{D}(c)} \left( \hat\tau_{D}(c) - \tau_{D}(c) \right)^{2} - \frac1{\tau_{D}(c) \hat\tau_{D}(c)} \left(\hat\tau_{Y}(c) - \tau_{Y}(c) \right) \left(\hat\tau_{D}(c) - \tau_{D}(c) \right).
\end{multline}
Since the treatment of the variance component is exactly the same as that by IK, we only discuss the bias component.
Observe that Lemma 1 of Arai and Ichimura (2015) implies the bias of $\hat\tau_{Y}(c)$ and $\hat\tau_{D}(c)$ are equal to
\[
C_{1}\left[ m_{Y+}^{(2)}(c) h_{+}^{2} - m_{Y-}^{(2)}(c)h_{-}^{2}\right] + \Bigl[\zeta_{Y1}(c) h_{+}^{3} - \zeta_{Y0}(c) h_{-}^{3}\Bigr] + o\left(h_{+}^{3} + h_{-}^{3}\right)
\]
and
\[
C_{1}\left[ m_{D+}^{(2)}(c) h_{+}^{2} - m_{D-}^{(2)}(c)h_{-}^{2}\right] + \Bigl[\zeta_{D1}(c) h_{+}^{3} - \zeta_{D0}(c) h_{-}^{3}\Bigr] + o\left(h_{+}^{3} + h_{-}^{3}\right),
\]
respectively.
Combining these with the expansion given by (\ref{eq:expansion}) produces the required result.

\section{Procedures to Obtain Pilot Estimates}
Procedures to obtain pilot estimates for $m_{Yj}^{(2)}(c)$, $m_{Yj}^{(3)}(c)$, $f(c)$, $f^{(1)}(c)$,  and $\sigma_{Yj}^{2}(c)$, for $j=+, -$, are exactly the same as those for the sharp RD design by AI (see Appendix A of AI).
Pilot estimates for $m_{Dj}^{(2)}(c)$, $m_{Dj}^{(3)}(c)$, $\sigma_{Dj}^{2}(c)$, for $j=+, -$, can be obtained by replacing the role of $Y$ by $D$ in Step 2 and 3 of Appendix A of AI. 
Pilot estimates for $\sigma_{YDj}(c)$, $j=+, -$ are obtained analogously to $\sigma_{Yj}^{2}(c)$.
We obtain a pilot estimate for $\tau_{D}(c)$ by applying the sharp RD framework with the outcome variable $D$ and the assignment variable $X$.

\section*{Reference}
\textsc{Arai, Y.}, and \textsc{H. Ichimura} (2015): ``Simultaneous selection of optimal bandwidths for the sharp regression discontinuity estimator,'' CIRJE-F-889, CIRJE, Faculty of Economics, University of Tokyo.\\
\textsc{Calonico, S.}, \textsc{M. D. Cattaneo}, and \textsc{R. Titiunik} (2014): ``Robust nonparametric bias-corrected inference in the regression discontinuity design,'' Econometrica, 6, 2295--2326.

\end{document}